\documentclass[prb,preprint]{revtex4-1} 

\usepackage{amsmath}  
\usepackage{amsfonts} 
\usepackage{graphicx} 
\usepackage{xcolor}
\usepackage[official]{eurosym}

\begin{document}

\title{A low-cost confocal microscope for the undergraduate lab}

\author{A. Reguilon}%
\author{W. Bethard}%
\author{E. Brekke}
\email{erik.brekke@snc.edu}

\affiliation{Department of Physics, St. Norbert College, De Pere, WI 54115}%

\begin{abstract}
We demonstrate a simple and cost-efficient scanning confocal microscope setup for use in advanced instructional physics laboratories. The setup is constructed from readily available commercial products, and the implementation of a 3D-printed flexure stage allows for further cost reduction and pedagogical opportunity.  Experiments exploring the thickness of a microscope slide and the surface of solid objects with height variation are presented as foundational components of undergraduate laboratory projects, and demonstrate the capabilities of a confocal microscope.  This system allows observation of key components of a confocal microscope, including depth perception and data acquisition via transverse scanning, making it an excellent pedagogical resource.
\end{abstract}

\maketitle


\section{\label{sec:intro}Introduction}

The design and use of optical instruments is an area of broad interest, appealing particularly to students at the intersection of biological fields with physics and engineering.  With the large number of undergraduate physicists interested in careers in biomedical fields, a confocal microscope provides an excellent opportunity to see how physics principles relate to state-of-the-art equipment used for image formation.  The properties, advantages, and applications of confocal micrsocopy have been carefully investigated in the biological community.\cite{Sheppard1997, Elliot2020, Erie2009} 

While many undergraduate optics courses spend significant time on the ideas of optical instruments and traditional microscopes, confocal microscopy is an important application that is often absent from the student experience.  There has been a variety of designs for homebuilt confocal microscopes in the lab, \cite{Xi2007, Hsu2017, Arunkarthick2014, Jennings2022, Shakhi2021, Gong2019} but these designs are often more focused on the production of an image than on student understanding of the physics underlying the operation of the instrument.  Previous versions were often either too expensive, or left essential elements hidden in commercial components, making them nonideal for undergraduate pedagogy.  

In this paper we present a simple scanning confocal microscope experiment, ideal for the undergraduate optics or advanced instructional lab environment.  Rather than being designed to minimize its size or acquire images on par with commercial confocal microscopes, our instructional setup is designed to clearly demonstrate the physics involved in the image acquisition method and to make the optical components easy to manipulate. Our setup provides an excellent system to illustrate scanning confocal microscopy and also gives experience with the calibration of a data acquisition process.

Our system is designed to be versatile, with potential to develop student experimental skills in the areas optical systems, electronics, and 3D printing. Projects can be adapted to suit the context of an advanced lab setting, an optics or electronics course, an independent student project, or a senior thesis.  In the following sections, we outline an experimental process that can be used as a whole or broken up into parts, depending on the pedagogical purposes of the instructor. In the first experiment, a microscope slide is used to illustrate the ability of a confocal setup to acquire depth information in a sample.  A second experiment involves using this setup to investigate height variations of a solid material.  Finally, either of these experiments can incorporate a 3D printed translation stage,\cite{Sharkey2016} which allows integration of programming and electronic control.

\section{\label{sec:design}Design}

\begin{figure}[htbp]
\centering
\includegraphics[width=15.5cm]{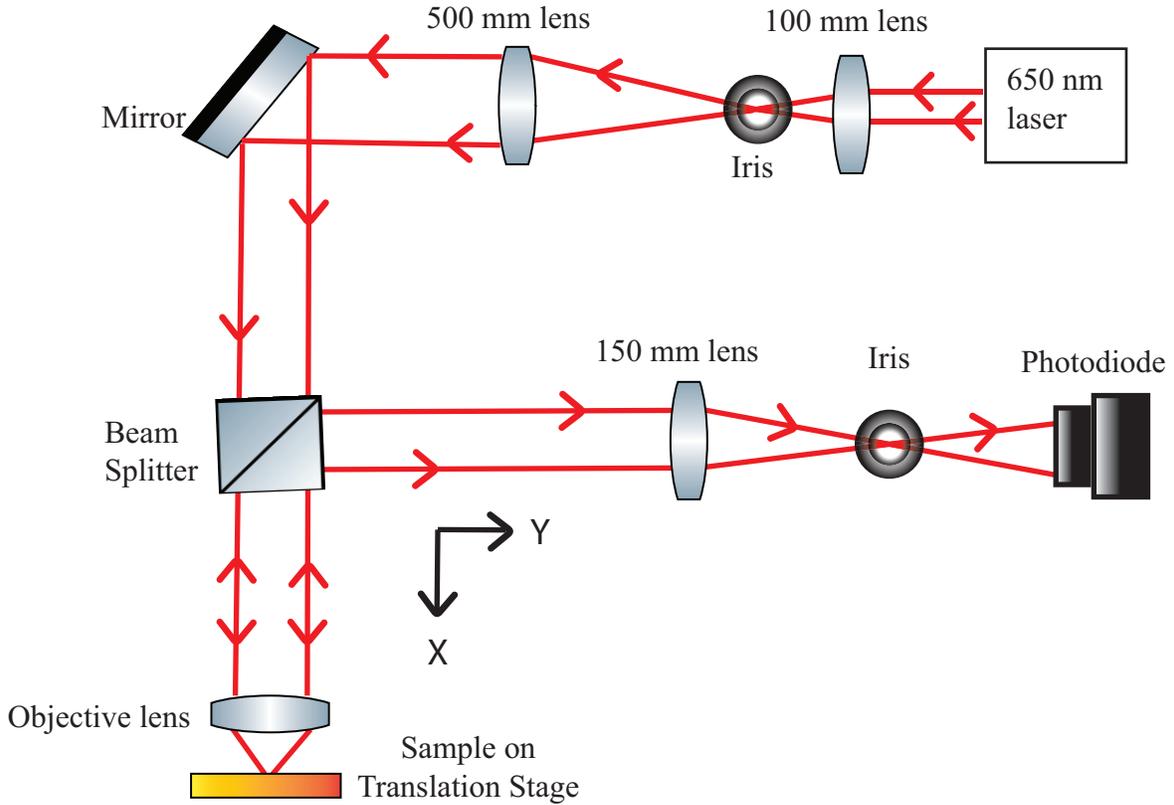}
\caption {The experimental setup for the confocal microscope. A generic visible laser is used, and expanded to fill a microscope objective lens.  The objective lens focuses the light onto a sample, which is mounted on a translation stage at the focus of the objective.  The light reflected from the sample is separated by a beam splitter, and focused onto an image plane, where an iris limits light to that which came from the focal plane of the objective.  This intensity is then monitored by a photodiode.}
\label{fig:model}
\end{figure}

Our confocal microscope experimental setup is shown in Fig.~\ref{fig:model}. The essential elements of this system are readily available commercial components. We use a simple 650 nm laser diode, with a optical power of 5 mW. This laser is expanded with a two-lens beam expander to a  beam waist of 5 mm waist to fill a commercial microscope objective lens.  The beam expander is constructed so that the distance between the lenses is the sum of the focal lengths, which increases the beam size by a factor of $f_2/f_1$. An iris is placed at the focus of the beam expander to make the beam more circular.  Once the beam is collimated, the distance between the expander and the objective, including the presence of a steering mirror, is chosen for convenience.  The sample is placed on a translation stage, which can be either a commercially purchased or a 3D-printed stage.  The objective focuses the light onto the sample, where a portion is reflected back through the objective and a beamsplitter redirects the light towards a photodetector.

\begin{table}[b]
\caption{\label{tab:table}
The parts required for construction of the confocal microscope.}
\begin{ruledtabular}
\begin{tabular}{cccccccc}
 Part&Specs&Supplier and part number&
 Price (US\textdollar)\\
\hline
Laser Diode & 650 nm, 5 mW & Adafuit 1054 & 10.00\\
Diode Holder &  & Thorlabs RA90 & 10.30\\ 
Feet (10) & 1$''$ x 2.3$''$ x 3/8$''$ & Thorlabs BA1S (5 PACK)&48.20\\
Lens Holder (7) & Ø1$''$ Optics, 8-32 Tap & Thorlabs LMR1 &109.83\\
Mirror Mount & Kinematic Mount for Ø1$''$ & Thorlabs KM100 & 39.86 \\
Mirror  &  Ø1$''$ 400 - 750 nm & Thorlabs BB1-E02 & 77.35\\
Objective Lens  & 10X infinite conjugate  & Amscope PL10X-INF-V300 & 66.99\\
Thread Adaptor  & SM1 to RMS   & Thorlabs SM1A3 & 18.50\\
Posts (10)  & 2$''$ & Thorlabs TR2-P5 & 50.52\\ 
Post Holders (10)  & 2$''$ & Thorlabs PH2-P5 & 81.28\\ 
150 mm Lens & Ø1$''$,  AR Coating: 650--1050 nm & Thorlabs LB1437-B &35.50\\
500 mm Lens &  Ø1$''$, AR Coating: 650--1050 nm & Thorlabs LA1908-B & 33.01 \\
100 mm Lens & Ø1$''$, AR Coating: 650--1050 nm & Thorlabs LA1509-B &34.39 \\
Beamsplitter & Economy Ø1$''$ 50:50 & Thorlabs EBS1 & 35.78\\
Ring Actuated Iris (2) & (Ø0.8 - Ø12 mm) & Thorlabs SM1D12D & 145.86 \\
Photodetector & 350 - 1100 nm & Thorlabs DET36A2 & 134.20 \\
Ultralight Breadboard & 24$''$ x 24$''$ x 0.98$''$ & Thorlabs PBG2424F & 842.93 \\
3-Axis RollerBlock & Long-Travel Bearing Stage & Thorlabs RB13M & 1,566.15 \\
\hline
&  & Total & 3,340.65  \\
\end{tabular}
\end{ruledtabular}
\end{table}

If the sample is at the focal point of the objective, the light coming back through the system will be collimated, and a final imaging lens will focus it through an iris before the photodetector.  Hence, if the sample is at exactly the focal length away from the objective, the image plane will be at the iris location, and a large portion of the light will make it to the photodetector.  However, if the sample's surface is not at the focal plane, the amount of light at the photodetector will be greatly reduced.  

Essential to the operation of the confocal microscope is the fact that the iris is in the image plane at the light focus.  Thus, it effectively images only the part of the sample at the focal plane of the objective lens.  Unlike traditional microscopes, this allows depth information in a sample.  

The apparatus described here is a cost-effective means for demonstrating these ideas in an undergraduate setting.  A full list of the parts needed can be seen in Table~\ref{tab:table}, where it is assumed power supplies and an oscilloscope are available.  The inclusion of the mirror and lens focal lengths were chosen by convenience with an available breadboard, and can be modified as desired.  The total cost of the equipment comes to  US  \$3,340.65.  However, the largest cost is the 3-axis translation stage, which can be replaced with a 3D-printed flexure stage,\cite{Meng2020, Sharkey2016} reducing the cost to under US  \$2,000 and opening up new opportunities for development.

This setup can be used to examine several simple objects, which provide excellent insight into the confocal imaging process. A deeper understanding of depth information from the microscope can come from examining standard microscope slides, especially those with concave sample openings.  The slide can be mounted on the translation stage in a number of ways; in our experiment, it was clipped to a 3D-printed stand, such that the laser passed through the slide with nothing behind it.  The system can also be used to examine surface variation of solid objects.  We have found that 3D-printed objects with known height variation works well, but a variety of other objects can be used.

\section{\label{sec:testing}Results and Analysis}

\begin{figure}[htbp]
\centering
\includegraphics[width=15cm]{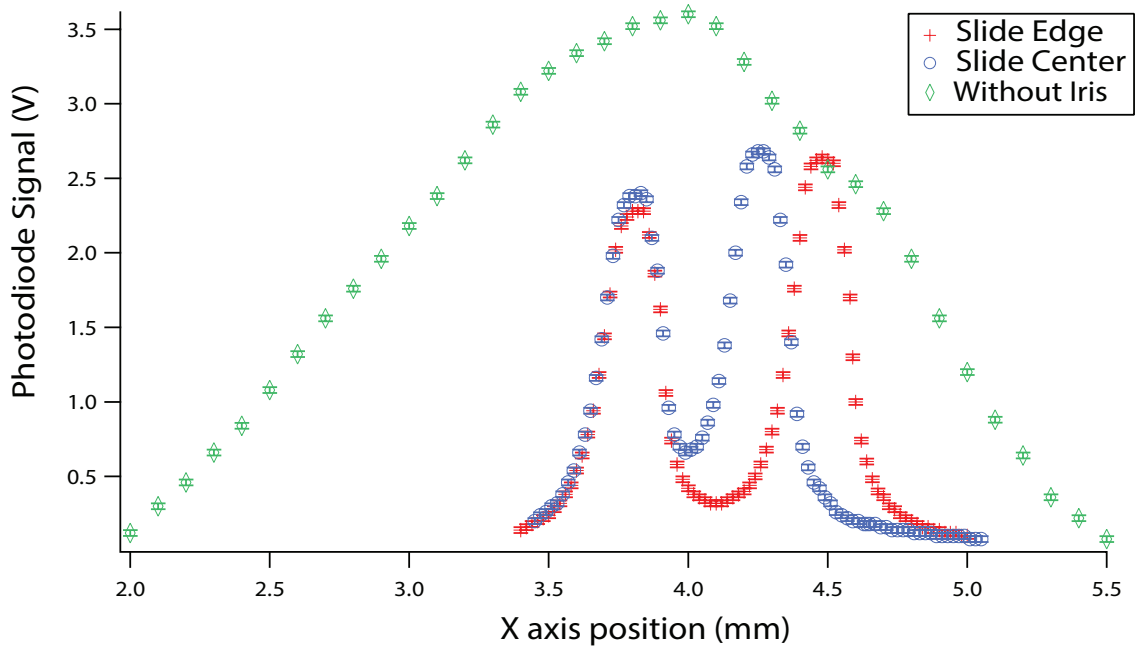}
\caption {As the distance between the objective lens and the slide is varied, intensity peaks are seen when either the front or the back surface of the slide is at the focal plane of the objective.   This is shown both at the flat edge of the slide, and at the curved center where the slide is not as thick.  The shifting of the second peak reveals the changing thickness of the slide.  Removing the iris in the imaging plane reveals the necessity of limiting the light from a single image plane for depth perception.}
\label{fig:signalexample}
\end{figure}

 In order to demonstrate the capabilities of the confocal microscope, two experiments are outlined here.  The first involves observing the thickness of semitransparent materials, with commercial microscope slides offering an excellent starting point.  Using a microscope slide to demonstrate how the confocal microscope works offers students the chance to determine depth information on a sample.  As a starting point, a microscope slide with a concave sample opening can be used.  In this case two reflections occur, one off of the slide's front surface and the other off of its back surface.  As a result, adjusting the position of the sample longitudinally along the direction of laser propagation allows the observation of two successive peaks, as each of the slide's front surface and back surface is in the focal plane of the objective.  An example of the data observed as the slide platform is adjusted longitudinally (in the $x$-direction) is shown in Fig.~\ref{fig:signalexample}. 
 Opening the iris at the image plane makes clear the way in which the confocal microscope causes a specific image plane for different depths in the material.  Without this iris, depth information is lost, as can be seen in Fig.~\ref{fig:signalexample}.

\begin{figure}[htbp]
\centering
\includegraphics[width=15cm]{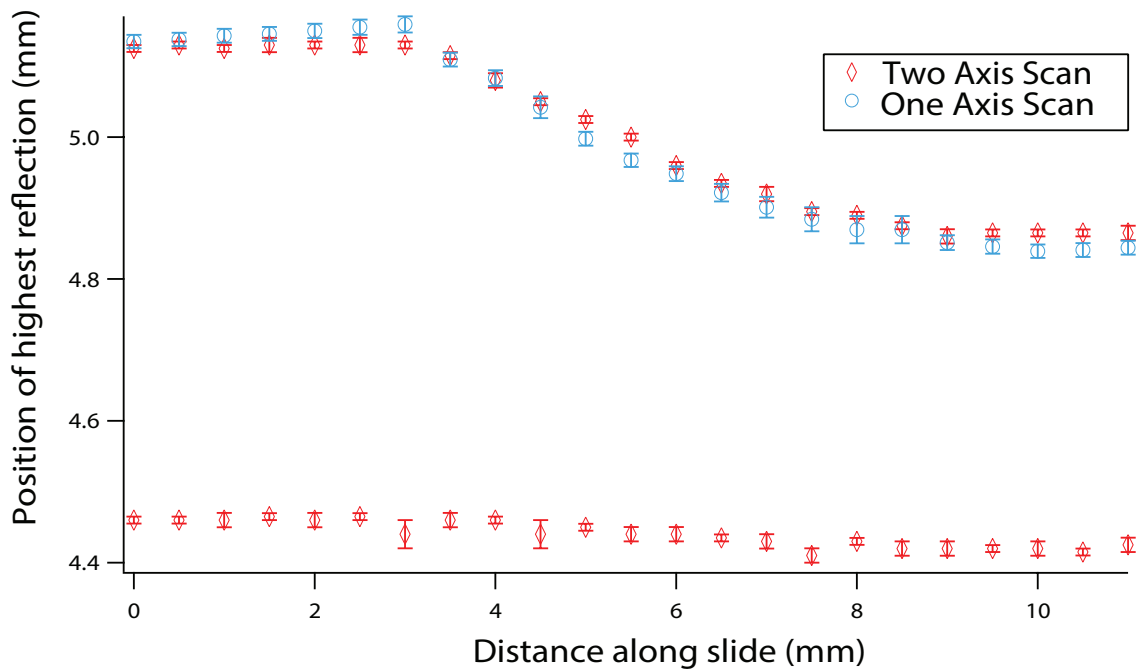}
\caption {The distance from the objective where peak reflected light is observed as a function of distance along the slide.  The region where the slide is flat is visible, before the concave opening begins at 4 mm. Then the changing thickness where the concave sample opening begins can be observed. The two-axis scan involves moving the translation stage transversely a small amount before scanning longitudinally to find the maximum.  The one-axis method involves only scanning transversely, using a known calibration of the photodiode voltage with depth. }
\label{fig:Curveslide.eps}
\end{figure}

By using a microscope slide with a concave sample opening, the thickness of the glass can be observed as a function of the position along the slide. As a starting point, this can be done by moving the slide a small distance transversely (in the $y$-direction), then scanning longitudinally (in the $x$-direction) to observe the two peaks, stepping through positions.  An example of the data obtained with this two axis scan technique is shown in Fig.~\ref{fig:Curveslide.eps}.

\begin{figure}[htbp]
\centering
\includegraphics[width=15.5cm]{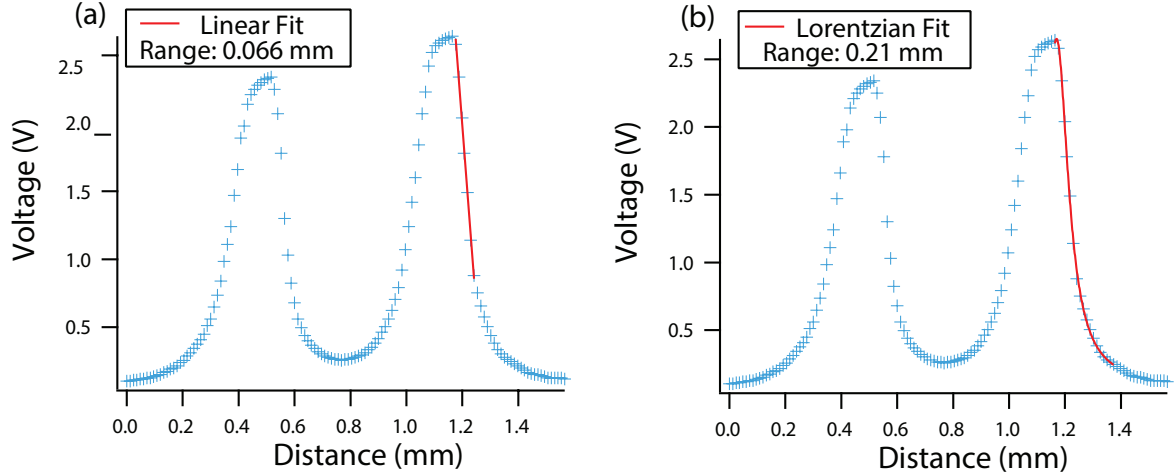}
\color{black} \caption { \color{black} The intensity peak obtained as the slide is translated longitudinally can be used to provide a voltage vs. height calibration for the sample.  An appropriate fitting function can be chosen to fit the peak over a particular height variation region on a sample.  a) A linear fit to a portion of the peak, providing a simple calibration valid over 0.066 mm around the sample.  b) A Lorentzian fit valid over 0.21 mm range around the sample. }
\label{fig:2fits}
\end{figure}

 This setup can also be used to demonstrate the usefulness of a scanning confocal microscope by acquiring the same depth information, while only moving the slide transversely to the laser beam.  This makes the image collection process much faster, but requires calibrating the system. As the slide is shifted transversely in the area where the depth varies, the different depths will cause different voltages in the photodiode. To correlate these voltage changes to height changes, a fit for the voltage as a function of height in the region scanned is found from the longitudinal scan, such as shown in Fig.~\ref{fig:2fits}.  
 
\color{black}

 The shape of the intensity peak is caused by the changing beam waist as it propagates through the image plane, with the intensity limited by the iris size.  This is a complex dependence, but can be roughly fit using a Lorentzian over much of the peak.  For a more precise calibration, it is desired to find a function that fits the intensity peak very well over a particular range of heights of the sample.  Using this calibration, voltage changes when scanning transversely can be used to calculate height changes in the sample.  The simplest calibration is to use a portion of the graph that can be well approximated as linear, as shown in Fig.~\ref{fig:2fits}(a).  However, this linear fit is only valid over a limited height variation, less than 100 $\mu$m.  Over a larger height variation, an exponential or polynomial fit may also work well, where we illustrate a Lorentzian fit in Fig.~\ref{fig:2fits}(b).  Allowing students to determine this calibration technique provides additional insight and experience in data analysis methods.  Taking the data again for the slide thickness using a transverse scan calibrated with the Lorentzian fit is shown in Fig.~\ref{fig:Curveslide.eps}, and agrees with that taken when collecting points individually with motion in both dimensions.  Further, this process allows students to understand how data can be collected and analyzed quickly in this `scanning' configuration of a confocal microscope.   
 
The ability of the confocal microscope to determine height variations of a solid sample represents another key category of experiment that can be undertaken.  To illustrate this point, stereolithography (SLA) 3D-printed objects of known height variation are used.  SLA 3D printing, like fused deposition modeling (FDM) 3D printing, creates a model layer by layer. Instead of extruding filament, a laser polymerizes resin at a specific point. This process allows for a quantifiable amount of material to be deposited with a known volumetric resolution. This property was exploited to create steps, with a known thickness, on the surface of a 3D-printed block small enough to demonstrate the accuracy of the system.  Variations on this experiment could be done with a variety of objects, as long as the object has height variation on the scale of 100s of microns.  

\begin{figure}[htbp]
\centering
\includegraphics[width=15cm]{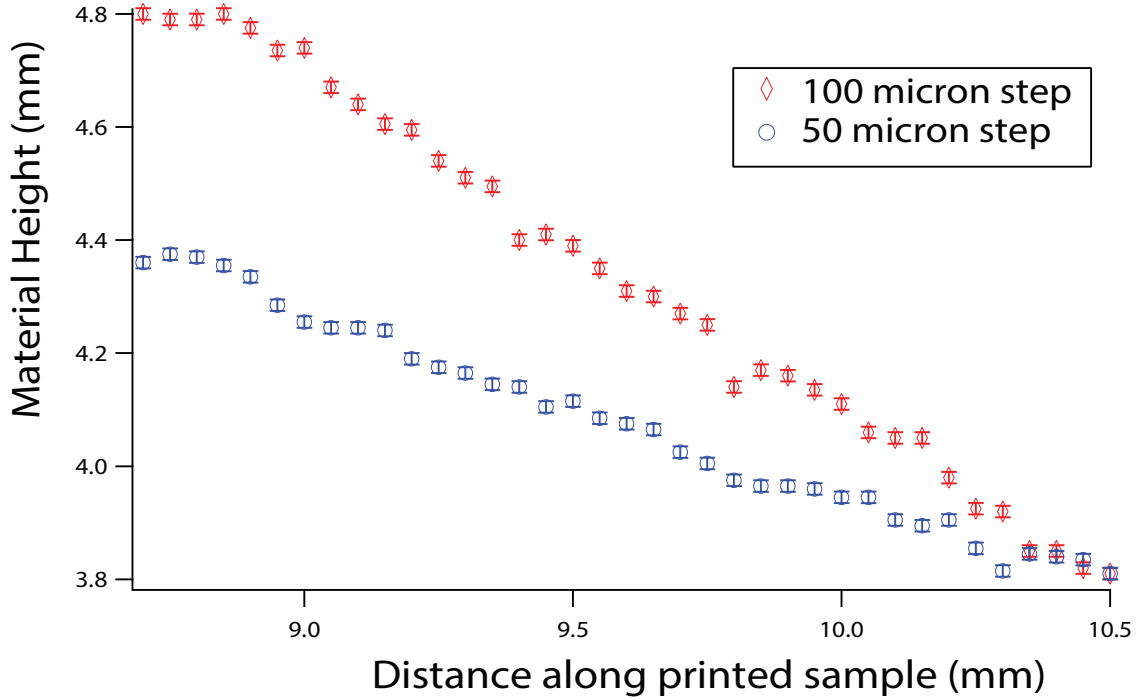}
\color{black} \caption { \color{black} The height of a solid 3D-printed object, as measured with the confocal microscope.  This object was designed and printed to show the capabilities of the microscope, and a number of similar objects can be used. }
\label{fig:solidsurface}
\end{figure}

With these solid objects, there are no longer multiple intensity peaks from different depths of the sample.  Instead, the location of the surface can be seen by the maximum of the intensity peak. Here, the surface height variation can be investigated either by successively stepping transversely and longitudinally, or by scanning only transversely using a calibrated intensity vs. height graph, as discussed above.  For our experiment two different surface variations were printed, one that increased 50 $\mu$m in height for each 200 $\mu$m shift in position, and another that increased 100 $\mu$m in height for each 200 $\mu$m shift in position.  Figure~\ref{fig:solidsurface} shows the measured surface location as a function of the distance along the sample for each of these materials using the two-axis scan technique, and illustrates the ability of the system to determine solid surface height variations.  The precision of 3D printers can limit the resolution of features created using this method, but it is expected that the cost of high-resolution 3D printers will continue to decrease, allowing more intricate surface patterns to be investigated.  Additional solid objects such as coins can also be examined, but the high reflectivity and angled surfaces can cause complex behavior in the signal intensity, and extra care is required.

\color{black}

 \section{\label{sec:3D print}Incorporation of 3D-printed translation stage}
 3D printing has developed into an extremely valuable tool for the undergraduate lab, and has been incorporated into a number of experimental designs.\cite{Mantia2022, Brekke2020, Schmidt2022}  Due to recent advances and open-source development, it is now possible to print and program excellent open-source translation stages,\cite{Sharkey2016, Meng2020} and 3D printing has been incorporated into a variety of microscope designs.\cite{Matsui2022, Collins2022, Diedrich2020, Rosario2022, Chagas2017, Brown2019}  In the confocal microscopy setup presented here, 3D printing presents the possibility of replacing a commercial translation stage system with a homebuilt option.  This replacement serves two main purposes: First, it can reduce the cost of the necessary equipment by a significant fraction. Second, it allows meaningful student experience in design, 3D printing, electronics, and programming.  This allows the confocal microscope project to be quite general, appealing to students interested in a wide range of engineering or data acquisition fields.  
 
 One option that incorporates 3-axis translation and excellent precision is the OpenFlexure Block stage. \cite{Meng2020} This setup is well documented elsewhere, and here was incorporated with little modification to the design presented.  This stage was specified to have a step size in the  \emph{x} direction of 12.4 $\pm$ 0.2 nm  and translate 2 mm on any axis.  While this translation distance is not large enough to scan over large portions of the slide, it does allow the same experimentation on slide thickness described above, with the data taken again as observed in Fig.~\ref{fig:signalexample}.  Having both the data for slide thickness taken from the commercial translations stage and the OpenFluxure Block stage allows an additional means of verifying the calibration for distance per step on the block stage.  Using these data, our calibration was determined to be 12.1 $\pm$ 0.4 nm, consistent with previous measurements.\cite{Meng2020}  The use of 3D-printed translation stages makes the scanning confocal microscope even more accessible, and would combine well with projects commonly undertaken in an undergraduate lab environment.

\section{\label{sec:future}Conclusions and Future directions}

The design for a homebuilt scanning confocal microscope presented here is ideal for an undergraduate instructional laboratory setting.  Through the measurement of microscope slide thicknesses and surface variations, undergraduates develop a better sense of the methods of scanning confocal microscopy and the information it provides.  This design provides a robust setup with understandable mechanisms at a low price that enables easy implementation.  

Our setup is intended for pedagogical purposes rather than quality imaging competitive with commercial confocal microscopes, but can be extended for more complete imaging.  If desired, the OpenFlexure stage can be automated, allowing for expansion of this project into acquiring scanned three dimensional images.  This is especially appealing in the context of an independent project or senior thesis looking to incorporate additional programming.   

An investigation of the resolution of the system\cite{Elliot2020} can also be employed as an extension to the experiments presented here.  
Investigating the limiting resolution would require materials with much finer features, as with this wavelength and numerical aperture lens the theoretical lateral resolution is expected to be near 1 $\mu$m, and the axial resolution on order 15 $\mu$m.  Here several factors could be investigated to determine maximum resolution, including beam size, the minimal size and location of the iris, and the numerical aperture of the objective lens used.      

The confocal microscope system described in this paper can easily be expanded, either to improve its quality or examine further applications.  Enhancements include the addition of an additional beamsplitter and camera, the use of achromatic doublets to decrease aberrations, improving the laser quality through fiber-coupling before use, or a galvo-scanning mirror in place of the $x$-$y$ translation stage.  In addition, the experiments outlined here can easily be extended to simple biological samples or fluorescence microscopy, allowing further insight into the way the optical system is commonly applied in biomedical fields.  

\begin{acknowledgments}
We wish to acknowledge the contributions of Joseph Coonen and the programming insight of Michael Olson.
\end{acknowledgments}





\end{document}